\documentclass[prl,twocolumn]{revtex4}
\usepackage{epsfig}
\begin{document}
\flushbottom

\title{Evidence for higher order QED in $e^+ e^-$ pair production at RHIC}
\author{A. J. Baltz}
\affiliation{Physics Department,
Brookhaven National Laboratory,
Upton, New York 11973}
\date{October 25, 2007}

\begin{abstract}
A new lowest order QED calculation for RHIC $e^+ e^-$ pair production
has been carried out with a phenomenological treatment of the Coulomb
dissociation of the heavy ion nuclei observed in the STAR ZDC triggers.
The lowest order QED result for the experimental acceptance is nearly
two standard deviations larger than the
STAR data.  A corresponding higher order QED calculation is
consistent with the data.
\\
{\bf PACS: { 25.75.-q, 34.90.+q}}
\end{abstract}
\maketitle

In a recent publication data were presented by the RHIC STAR collaboration
on ``Production of $e^+ e^-$ pairs accompanied by nuclear dissociation in
ultraperipheral heavy-ion collisions''\cite{star04}.  In this publication
the authors compared the experimental cross section $\sigma$ with a lowest
order QED calculation $\sigma_{QED}$ and concluded that ``At a 90\%
confidence level, higher-order corrections to the cross section,
$\Delta \sigma = \sigma -\sigma_{QED}$ must be within the range
$-0.5 \sigma_{QED} < \Delta \sigma < 0.2 \sigma_{QED}$''.

In this Letter, I will present a new lowest order QED cross section calculation
result significantly larger than that presented in the STAR publication.  This
new lowest order QED cross section is now two standard deviations
above the STAR experimental result.  Furthermore, I find that a higher
order QED calculation gives good agreement with the experimental result.

The STAR experimental cross sections are for a final state with an
$e^+ e^-$ pair in combination with mutual nuclear excitation of both ions
by the ions' Coulomb fields observed in the Zero Degree Calorimeters (ZDC).
Thus impact parameter dependent probabilities have to be calculated
both for $e^+ e^-$ pair production and for mutual Coulomb dissociation
of the ions.

The calculations presented here make use of the method of calculating
the impact parameter dependent probability of producing $e^+ e^-$ pairs
previously described\cite{ajb06}.  Use of an appropriate gauge allowed
the electromagnetic fields of colliding ions to be represented as colliding
delta functions in the ultrarelativisitic limit.  The solution of the 
time-dependent Dirac equation for lepton pair production was thereby obtained
in closed form.  This solution contains higher order QED effects and goes
to the known perturbative (lowest order) result in that limit.  It is well
known that photoproduction of $e^+ e^-$ pairs on a heavy target shows a
negative (Coulomb) correction proportional to $Z^2$ that is well described
by the Bethe-Maximon theory\cite{bem}.  The calculated higher order
heavy ion pair production cross sections show analogous reductions
from perturbation theory.

The impact parameter ($b$) dependent amplitude presents a particular numerical
challenge since it involves the a rapidly oscillating phase
$\exp(i \bf{k \cdot b})$ in the integral over the transverse momentum $\bf{k}$
transfered from the ion to the lepton pair.  The usual method of evaluating
the perturbative impact parameter dependent probability is to first square
the amplitude and then integrate over the sum and difference of
$\bf{k}$ and $\bf{k^\prime}$.  Here I have integrated before squaring, and
I deal with the rapid oscillations with the piecewise analytical integration
method previously described\cite{ajb06}, in particular Eq (A5-A7).
In that previous $b$ dependent calculation of the total cross section, half of
the contribution comes from $b > 5000$ fm and contributions up to $b = 10^6$
fm are considered.  Due to the large values of $b$ contributing, that
calculation was somewhat crude.  However integration over $b$ reproduced the
known cross sections calculated with the $b$ independent method or calculated
from the very accurate analytical Racah formula\cite{rac} to about 3\%.
It can also be noted that the computed perturbative and eikonal $b$ dependent
probabilities in that paper were in relatively good agreement with
calculations in the literature\cite{hbt1} available for $ b < 7000 $ fm.

In the present calculation although $\bf{k}$ takes on large values scaled
to the momenta of the produced pairs, the range of of $b$ considered is
limited to under 100 fm by the mutual Coulomb dissociation requirement of
the ZDC tagging.  The calculated high momentum piece
of the cross section was taken as described by the STAR
experimentalists\cite{star04,mrz}.  Electron and positron tracks were
required to have transverse momentum $p_T > 65$ MeV/c and pseudorapidity
$| \eta | < 1.15$.  In addition, at least one of the tracks was
required to have momentum $p < 130$ MeV/c.  Pairs were required to have
masses $140$ MeV $< M_{ee} < 265$ MeV, pair transverse momentum
$P_T < 100$ MeV/c, and pair rapidity $| Y | < 1.15$.
In the numerical calculations the mesh size
(piece) over which the analytical $\bf{k}$ integrations is taken was chosen
such that doubling the number of mesh points (halving piecewise size) changed
the computed result by less than 0.1\%.

Several exploratory calculations  were carried out to check the accuracy of
the lowest order $e^+ e^-$ calculations in the high momentum ranges here
considered.
The same expression for the pair production amplitude, derived in
the ultrarelativistic limit, is the basis for both impact parameter
independent\cite{ajb05} and dependent\cite{ajb06} cross section expressions.
The impact parameter independent calculation of the untagged
cross section in an approximate STAR acceptance range carried out by
Hencken, Baur, and Trautmann\cite{hbt}, 0.322 b, was reproduced with a
modified version of my impact parameter independent program\cite{ajb05},
and with a version of the code LPAIR\cite{boc,ver}, both to one percent
accuracy.

The nuclear breakup was calculated using a phenomenological model based on
photodissociation data as imput for Weizs\"{a}cker-Williams
calculations\cite{brw} for
mutual Coulomb dissociation\cite{bcw} that was found to successfully
reproduce cross sections for mutual Coulomb dissociation observed
in the zero degree calorimeters at RHIC\cite{chiu}.  This phenomenological
approach to the nuclear breakup was further incorporated in
successful calculations of tagged vector meson production probabilities
at RHIC\cite{bkn,star02}.
 
In analogy with the previous tagged vector meson calculations\cite{bkn}
and STAR equivalent photon calculations\cite{star04},
the cross section here is computed from the product of the pair production
probability $P_{ee}(b)$, the Coulomb dissociation probability $P_{xx}(b)$,
and a factor $\exp[-P_{nn}(b)]$ to exclude events where hadronic interactions
occur
\begin{equation}
\sigma = 2 \pi \int P_{xx}(b) P_{ee}(b) \exp[-P_{nn}(b)] b db .
\end{equation}
$P_{xx}(b)$ is the unitarized probability that both colliding nuclei suffer
Coulomb dissociation
\begin{equation}
P_{xx}(b) = [1 - \exp(-P_C(b)]^2
\end{equation}
with $P_C(b)$ the non-unitarized probability of a single Coulomb excitation.
The hadronic interaction probability $P_{nn}(b)$ is calculated in the usual
Glauber manner from impact parameter dependent nuclear density (with
Woods-Saxon shape R=6.38 fm
a=.535 for Au) overlap and a nucleon-nucleon cross section of 52 mb for
the RHIC center of mass energy of 200 GeV.  In calculating
the pair production probability $P_{ee}(b)$
an analytical elastic form factor was employed
\begin{equation}
F(q)={3 \over (q r)^3} [ \sin(q r) - q r \cos(q r) ]
\Biggl[{ 1 \over 1 + a^2 r^2} \Biggr]
\end{equation}
with a hard sphere radius $r = 6.5$ fm and Yukawa potential range
$a = 0.7$ fm.  This form very closely reproduces the Fourier transformation
of the Au density with the Woods-Saxon parameters 
mentioned above\cite{kn}.
Test calculations with an alternative form factor that has been utilized in
the literature\cite{kai2} for analytical simplicity, the dipole form with
$\Lambda=83$ MeV/c, gave results in agreement with those making use
of the analytical Woods-Saxon form to one percent.

The lowest order QED calculation (1.9 mb) used by STAR for comparison with
data (1.6 mb)
mentioned above made use of an approximate analytical treatment of the Coulomb
dissociation\cite{hbt} combined with lowest order QED for the pair production.
An alternate lowest order calculation (2.1 mb) presented by
the STAR authors made use of the realistic Coulomb dissociation
phenomenology\cite{brw,bcw,bkn} observed in the ZDC triggers such as adopted in
the calculations here but with an equivalent photon QED calculation of the
$e^+ e^-$ pairs.   Both these calculations
left something to be desired.  The lowest order QED calculation,
made use of an overly simplified approximation to the Coulomb dissociation
phenomenology and involved choosing an arbitrary lower cutoff for the impact
parameter integration.  The equivalent photon calculation, did not include
photon virtuality.    The QED calculations presented
here, both lowest order and higher order, have neither of these shortcomings.

It is useful to change variables for the impact parameter integration
in order that the contribution to the cross section is a smooth function
of a variable that goes to zero at the high and low end.  Choose
a new variable $u = (b_1/b)^{3/2}$
and for convenience set $b_1$ equal to 13 fm.  Figure 1 shows the
contribution to the cross section as a function of $u$. 
\begin{figure}[h]
\begin{center}
\epsfig{file=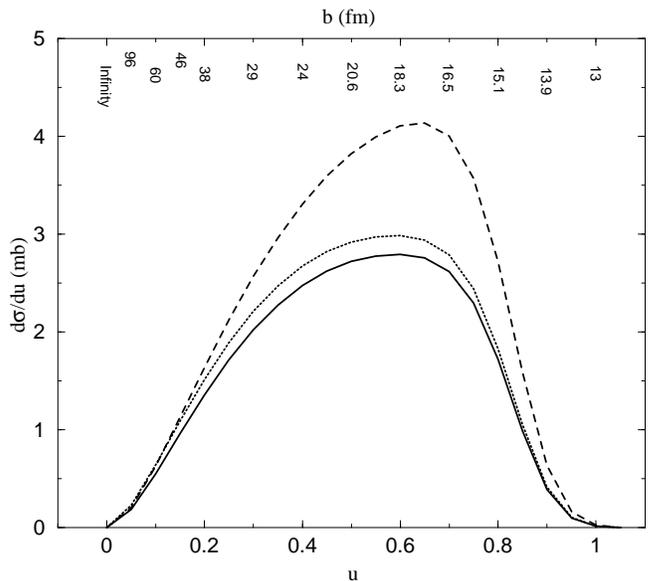,height=8cm}
\end{center}
\caption{Tagged pair production cross section as a function of $u$.
Corresponding values of the impact parameter $b$ are indicated next to tick
marks at the top.  The dashed line is for the lowest order QED calculation,
the solid line for the higher order QED calculation, and the dotted line for
an eikonal higher order QED calculation.}
\end{figure}
It is interesting to note that although the original simple eikonal inclusion
of higher order effects gives the same result as perturbation theory for
untagged pair production, the weighting of low impact parameter pairs
due to the Coulomb dissociation of the STAR acceptance lead to an
eikonal higher order result that differs from perturbation theory.  Between
$b=46$ and $b=60$ the eikonal result
crosses over from being below the perturbation theory result at low $b$ to
above the perturbation theory result at high $b$.  This crossover has been
noted at about 3000 fm in numerical calculations of the total $e^+ e^-$
pair production probability at RHIC\cite{hbt1,ajb06}.

The integrated lowest order QED cross section is 2.34 mb for the STAR
acceptance.  This is nearly two standard deviations higher than the STAR
experimental result of $1.65\pm 0.23$(stat)$\pm 0.30$(syst) mb\cite{mrz}.  
The corresponding higher order QED calculation result of 1.67 mb is
consistent with the STAR data.  Thus the STAR data seem to provide some
evidence of higher order QED in ultrarelativistic heavy ion reactions.
The integrated eikonal result, 1.80 mb, is also significantly reduced from the
lowest order perturbative result.

I would emphasize that the higher order QED result here was calculated with the
same physical cutoff of the transverse potential that restored the
Coulomb corrections to calculations of the total cross section\cite{ajb05},
qualitatively consistent with the approximate higher order
calculations carried out by Ivanov, Schiller and Serbo\cite{serb} and
by Lee and Milstein\cite{lm1,lm2}.  In particular, the higher order QED effect
from the physical cutoff was shown numerically to go over to the Coulomb
correction result of Lee and Milstein in their small $\bf{k}$
expansion limit\cite{ajb3}.  The eikonal result makes use
of the original formulation
of Baltz and McLerran\cite{bmc} and of Segev and Wells\cite{sw1,sw2}
(first calculated numerically by Hencken, Trautmann, and Baur\cite{hbt1})
that was not consistent with the more proper regularization of Lee and
Milstein.

\begin{figure}[h]
\begin{center}
\epsfig{file=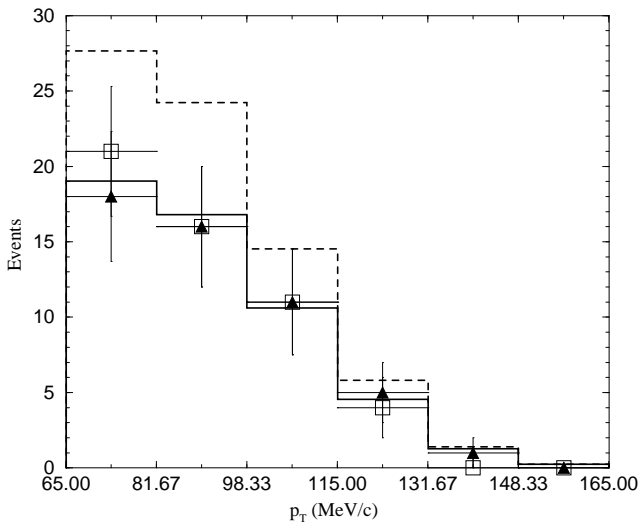,height=7cm}
\end{center}
\caption{$p_T$ spectra for the produced positrons and electrons. Squares are
the number of positrons and triangles electrons.  The dashed line is the lowest
order QED calculation and the solid line higher order QED.}
\end{figure}
Figure 2 is a comparison of the $p_T$ spectra for the number of produced
positrons and electrons.  A total of 52 events was observed in the STAR
acceptance.
The dashed line represents the number of expected positrons
or electrons within each bin calculated in lowest order QED.  The solid line
represents the corresponding expected numbers from the higher order QED
calculation.  Obviously the higher order calculation is in good agreement
with the data, while the lowest order calculation overpredicts.

\begin{figure}[h]
\begin{center}
\epsfig{file=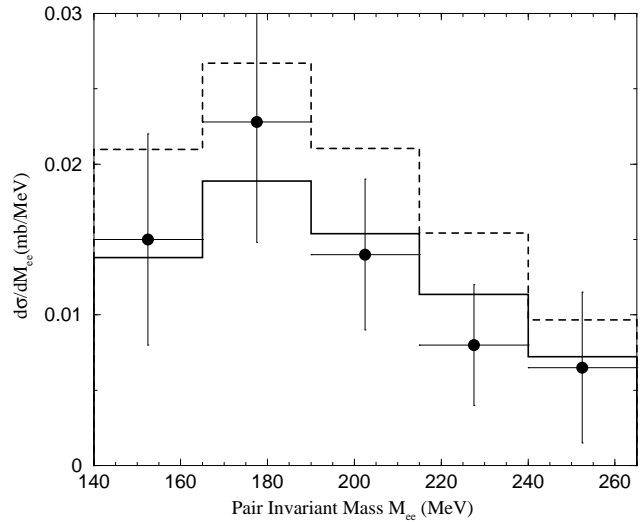,height=7cm}
\end{center}
\caption{The pair equivalent mass distribution.}
\end{figure}
Figure 3 shows pairs cross section data (filled circles) as a function of
the pairs
invariant mass to be more consistent with the solid line higher order QED
calculation than with the dashed line lowest order QED.  Likewise the
data shown as a function of the pair momentum in Figure 4 show good
agreement with the higher order QED calculation but are below the lowest
order calculation.

\begin{figure}[h]
\begin{center}
\epsfig{file=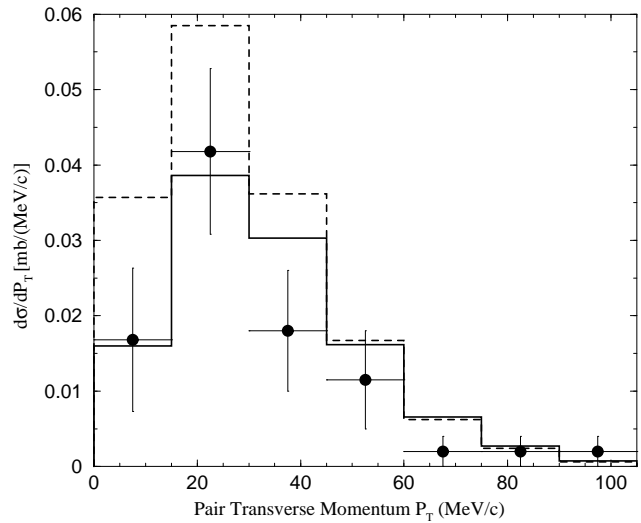,height=7cm}
\end{center}
\caption{The pair $P_T$ spectrum.}
\end{figure}

In these calculations it has been as crucial to get the overall normalization
of the lowest order QED calculation correct as to calculate the reduction
due to higher order effects.  This normalization depended on reliable
QED pair production calculations, reliable Coulomb dissociation calculations,
and faithful reproduction of the STAR experimental momentum cuts.
But if these calculations are accepted as reliable, then the above comparison
of the QED calculations with the STAR data provides the first evidence of
higher order QED in relativistic heavy ion reactions.

In the planning stages of RHIC a workshop was held at
Brookhaven on the topic ``Can RHIC be used to test QED?''\cite{frt}  A recent
paper concluded, ``We think that after about 17 years the answer to this
question is 'no'.''\cite{bht07}  The present results indicate that the
answer may turn out to be 'yes'. 

I would like to acknowledge helpful communications with Spencer Klein and
Dariusz Bocian. 

This manuscript has been authored
under Contract No. DE-AC02-98CH10886 with the U. S. Department of Energy.

\end{document}